\documentclass[twocolumn,floatfix,prl, aps,superscriptaddress]{revtex4-2} 
\usepackage{graphicx,amsfonts,amssymb,amsmath,hyperref,hypcap,enumerate,enumitem}
\usepackage[dvipsnames]{xcolor}
\usepackage{soul}
\usepackage{scalerel}
\usepackage{adjustbox,stmaryrd}
\usepackage{cancel}
\usepackage{csquotes}
\usepackage[utf8]{inputenc}
\usepackage{geometry}
\usepackage{physics}
\usepackage[normalem]{ulem} 

\hypersetup{  colorlinks=true, linkcolor=blue, citecolor=blue, urlcolor=blue  }

\begin{document}
\title{Crystallization in the Fractional Quantum Hall Regime with Disorder-Aware Neural Quantum States}
\author{Jihang Zhu}
\thanks{These authors contributed equally to this work and are considered co-first authors.}
\affiliation{Department of Materials Science and Engineering, University of Washington, Seattle, Washington 98195, USA}
\author{Yi Huang}
\thanks{These authors contributed equally to this work and are considered co-first authors.}
\affiliation{Department of Materials Science and Engineering, University of Washington, Seattle, Washington 98195, USA}
\author{Xiaodong Hu}
\affiliation{Department of Materials Science and Engineering, University of Washington, Seattle, Washington 98195, USA}
\author{Di Xiao}
\affiliation{Department of Materials Science and Engineering, University of Washington, Seattle, Washington 98195, USA}
\affiliation{Department of Physics, University of Washington, Seattle, WA, USA}
\author{Ting Cao}
\affiliation{Department of Materials Science and Engineering, University of Washington, Seattle, Washington 98195, USA}

\begin{abstract}
We present the first microscopic demonstration of a disorder-pinned hole Wigner crystal (WC), providing a natural explanation for the reentrant integer quantum Hall effect observed near $\nu=2/3$, as well as its analogs in fractional Chern insulators. We further identify a novel crossover regime above filling $\nu=2/3$ that connects this hole WC to an electron WC, characterized by a network-like electron density structure. To uncover these phenomena, we use neural-network variational Monte Carlo (NNVMC) with a disorder-aware self-attention neural quantum state that describes both fractional quantum Hall (FQH) liquids and Wigner crystals within a single unbiased variational framework.
More broadly, our method establishes a unified phase diagram that exposes a fundamental asymmetry in crystallization across half-filling:
near $\nu=1/3$, increasing LL mixing and disorder both stabilize an electron WC, whereas near $\nu=2/3$, the hole WC dominates at weak LL mixing and ultimately gives way to the electron WC at strong LL mixing.
\end{abstract}

\maketitle

{\em Introduction}---In a two-dimensional electron system (2DES) under a strong perpendicular magnetic field, incompressible fractional quantum Hall (FQH) liquids~\cite{Tsui1982} compete with charge-ordered Wigner crystals (WCs)~\cite{Wigner:1934,Lozovik:1975,Andrei:1988,Santos:1992,Young:2021}. Early theory for a clean 2DES, considering only Coulomb interactions projected to the lowest Landau level (LLL), predicted a liquid-solid transition upon decreasing the filling factor below $\nu_c \approx 1/6.5$~\cite{Girvin:1984}. 
Experiments, however, have revealed a much more intricate phase diagram where the evolution is not simply a monotonic termination of FQH liquids into a WC. 
Instead, reentrant insulating behavior can emerge below half-filling ($\nu < 1/2$), with robust FQH liquids flanked by insulating electron WC (e-WC) phases~\cite{Willett:1988,Goldman:1988,Jiang:1990,Jiang:1991,Sajoto:1990,Santos:1992,Du:1996,Pan:2002,Yazdani2024}; above half-filling ($\nu > 1/2$), a particle-hole symmetric reentrant integer quantum Hall (IQH) effect occurs~\cite{Eisenstein:2002,Li_short_range:2010}, associated with an insulating hole WC (h-WC)~\cite{Maki1983,Huang:2025} residing on an IQH background that maintains a quantized Hall conductivity $\sigma_{xy} = e^2/h$.
Recent experimental reports on the reentrant integer quantum anomalous Hall (IQAH) effect in fractional Chern insulators (FCIs)~\cite{Lu:2025,Xu:2025} have further renewed interest in this liquid-solid competition.

The complexity of this phase diagram arises because the liquid-solid phase boundary is highly sensitive to both Landau level (LL) mixing~\cite{Santos:1992,Santos:1992b,Jain:2018,Maryenko:2018,Wang:2025b,Yazdani2024} and short-range disorder~\cite{Santos:1992,Li_short_range:2010,Moon:2014}. In real materials, unavoidable disorder and material-dependent LL mixing strongly reshape the liquid-solid phase competition. While clean, low-LL-mixing GaAs electron systems restrict WCs to $\nu \lesssim 1/5$~\cite{Madathil:2024,Wang2025}, enhancing LL mixing (as in ZnO~\cite{Maryenko:2018}, GaAs hole systems~\cite{Santos:1992,Santos:1992b,Wang:2025b}, and graphene~\cite{Yazdani2024}) or introducing short-range alloy disorder~\cite{Li_short_range:2010,Moon:2014} pushes the crystalline boundary to significantly higher fillings near $\nu=1/3$ and $2/5$. Yet, despite the prominent role of these two ingredients in experiments, a unified treatment of LL mixing and disorder in the strongly correlated FQH regime remains exceptionally difficult.
Existing methods each face limitations. 
Exact diagonalization (ED) is limited to small system sizes and typically to the lowest few LLs, making strong LL mixing difficult to access~\cite{Haldane:1985}.
Variational and diffusion Monte Carlo (VMC and DMC) can incorporate LL mixing, but they rely on distinct trial wave functions for liquid and crystal phases and are sensitive to the choice of ansatz~\cite{Zhu1993,Zhu1995,Price1995,Ortiz1993,Jain:2018, MGattu:2025}.
Disorder further increases the complexity, and existing studies are largely restricted to small systems or simplified impurity models~\cite{Huang2025Thermal,Zhang1985,Rezayi1985,Jain:2007}.

In this work, we overcome these limitations using neural-network variational Monte Carlo (NNVMC)~\cite{GCarleo_NNVMC_2017,psiformer,YQian_FQH_NNVMC_2025,Teng:2025,Abouelkomsan:2026,nazaryan2026} with a disorder-conditioned neural quantum state (NQS), enabled by large-scale GPU computing. Our approach introduces two key advances. First, the NQS provides a single, highly expressive real-space ansatz that can represent both FQH liquids and WCs in the full Hilbert space with all LLs without truncation or phase-specific bias, allowing the ground state to be determined variationally. 
Second, we incorporate disorder directly into the neural-network architecture through impurity-aware features within a self-attention framework, enabling the ansatz to capture spatial inhomogeneity together with LL-mixing effects. This unified framework enables, to our knowledge, an unprecedented large-scale nonperturbative treatment of LL mixing and disorder in a single variational description.

\begin{figure}[t]
\centering
\includegraphics[width=\linewidth]{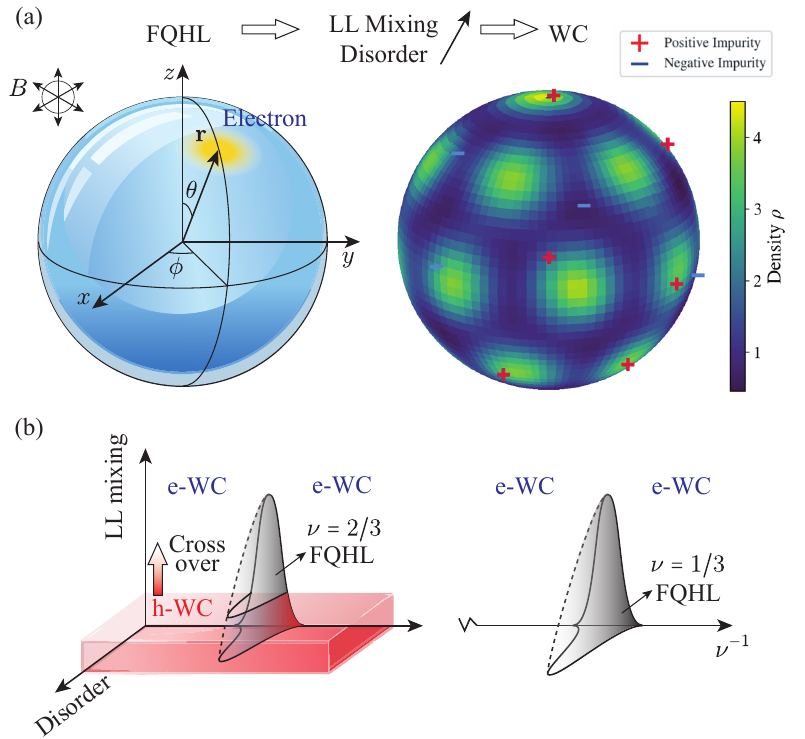}
    \caption{Schematics of liquid-solid phase competition in strong magnetic field. (a) Electrons on a Haldane sphere with a radial magnetic field $B$. Strong LL mixing or disorder destabilizes a uniform FQH liquid (FQHL, left) in favor of a charge-ordered WC pinned by positive and negative impurities (right). (b) Schematic 3D phase diagram in the parameter space of LL mixing, disorder, and $\nu^{-1}$. Near $\nu=1/3$, an e-WC phase surrounds the central FQHL dome. Near $\nu=2/3$, a h-WC emerges at weak LL mixing as the particle-hole counterpart to the e-WC near $\nu=1/3$. At intermediate LL mixing, a crossover region may connect the h-WC and e-WC phases. Generally, increasing disorder drives the crystallization of the FQHL into either an e-WC or h-WC. In the schematic, the LL-mixing axis starts from $\kappa=1$ and the disorder axis starts from two impurities, reflecting the parameter range accessed in our calculations.} 
\label{fig:illustration}
\end{figure}

\begin{figure}[t]
    \centering
    \includegraphics[width=\linewidth]{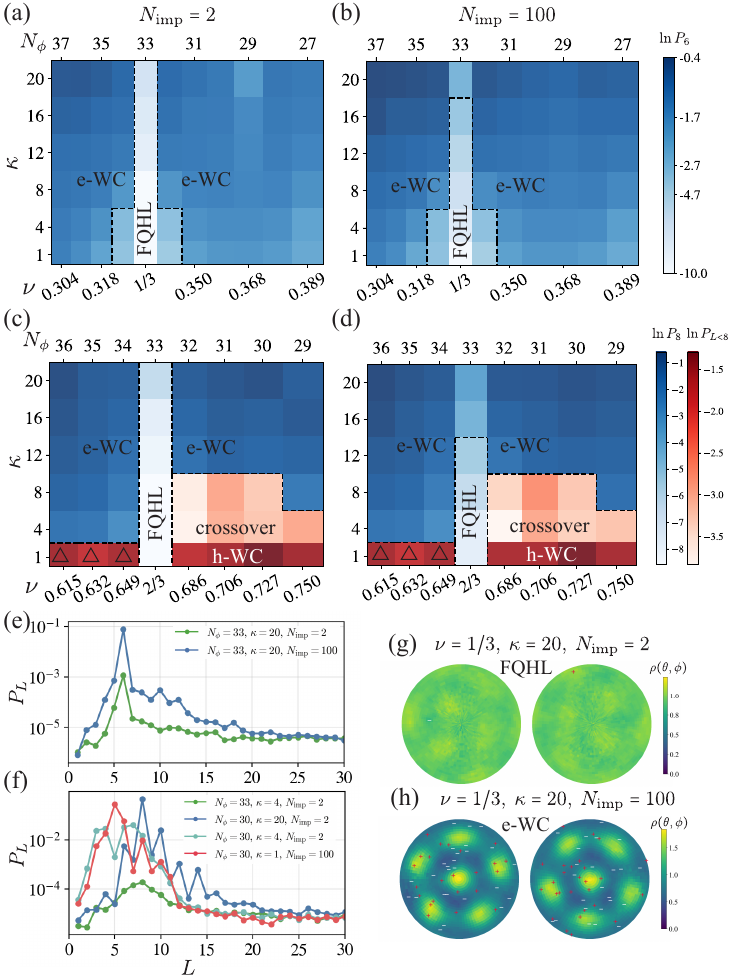}
    \caption{Phase diagrams for systems with $N_{\rm imp} = 2$ and $100$ impurities at fixed impurity strength $\kappa_{\rm imp}/\kappa = 0.01$. The filling factor $\nu$ is tuned by varying the flux $N_{\phi}$. (a,b) Phase diagrams near $\nu=1/3$ for $N=12$ electrons. Color scale represents the crystalline order parameter $\ln P_6$, and the dashed line at $\ln P_6 = -2.7$ approximates the boundary between WC and FQHL. (c,d) Phase diagrams near $\nu=2/3$ for $N=22$ electrons. Blue (red) color scale represents the electron (hole) crystalline order parameter $\ln P_8$ ($\ln P_{L<8}$). The WC-FQHL boundary is approximated by $\ln P_8 = -4.6$. 
    The triangles at $\kappa=1, \nu<2/3$ mark the cases not fully converged for $N=22$ electrons, but identified as an intermediate crystalline texture carrying signatures of both e-WC and h-WC order in its angular power spectrum from converged smaller-system results using $N=12$ electrons (see Supp).
    (e,f) Angular power spectra (e) at $\nu=1/3$ and (f) near $\nu=2/3$.
    (g,h) Electron density profiles projected onto the north and south hemispheres at $\nu=1/3$ and $\kappa=20$ for (g) $N_{\rm imp} = 2$ and (h) $N_{\rm imp} = 100$.}
    \label{fig:phase_diagram_results}
\end{figure}

Using this approach, we uncover qualitatively distinct crystallization behavior below and above $\nu=1/2$, establishing the schematic phase diagram in Fig.~\ref{fig:illustration} based on our quantitative results [Fig.~\ref{fig:phase_diagram_results}].
Exactly at $\nu=1/3$ and $\nu=2/3$, FQH liquid (FQHL) remains more robust against crystallization than nearby fillings.
Near $\nu=1/3$ [Fig.~\ref{fig:illustration}(b), right], increasing LL mixing $\kappa$ and the number of short-range impurities $N_{\rm imp}$ both stabilize an e-WC. 
Near $\nu=2/3$ [Fig.~\ref{fig:illustration}(b), left], we identity for an h-WC at weak LL mixing $\kappa$, marking its first demonstration via unbiased microscopic calculations.
This h-WC overtakes both the FQHL and the e-WC in the surrounding filling range. 
Since a disorder-pinned h-WC macroscopically manifests as an IQH plateau, its interruption by the $\nu=2/3$ FQHL provides a natural explanation for the reentrant IQH (IQAH) behavior observed in both GaAs and FCIs~\cite{Li_short_range:2010,Lu:2025,Xu:2025,Huang:2025}. 
We also discover an intermediate-$\kappa$ crossover regime above $\nu=2/3$ that connects the h-WC and e-WC, and is characterized by a network-like electron-density structure.
This regime reveals that particle-hole-related h-WC at weak $\kappa$ and Coulomb-driven e-WC at strong $\kappa$ may not be connected trivially. Because the h-WC carries $\sigma_{xy} = e^2/h$, whereas the e-WC is a trivial insulator, this crossover involves both a reorganization of crystalline order and a change in Hall response. A key question arising from our work is whether this regime corresponds to a first-order transition, a continuous transition, or a more intricate sequence of competing states.

{\em Hamiltonian and the wave-function ansatz}---We study $N$ interacting electrons on a spherical geometry (Haldane sphere) \cite{Wu:1976ge, Haldane_sphere_1983} in the presence of quenched short-range impurities. The full Hamiltonian is given by
\begin{equation}
\label{eq:hamil}
H = \sum_{i=1}^{N} \frac{|\pmb{\Lambda}_i|^2}{2 \kappa R^2} + \sum_{i<j} \frac{1}{|\pmb{r}_i - \pmb{r}_j|} + \frac{\kappa_{\rm imp}}{\kappa} \sum_{i=1}^{N} V_{\rm imp}(\pmb{r}_i).
\end{equation} 
$\pmb{\Lambda}_i$ is the dynamical angular momentum of the $i$-th electron~\cite{Haldane_sphere_1983}.
$N$ is the total number of electrons.
Throughout this work, lengths are in units of the magnetic length $\ell_B = \sqrt{\hbar c/eB}$, and energies in units of $e^2/\epsilon \ell_B$.
The LL-mixing strength is characterized by the ratio between Coulomb energy scale and the LL spacing $\kappa = e^2/(\epsilon \ell_B \hbar \omega_c)$.

The sphere is threaded by $N_{\phi} = 2Q$ flux quanta such that it has radius $R=\sqrt{Q}$. 
On a Haldane sphere, the filling factor is related to the flux by $2Q = N/\nu-\mathcal S$, with a Wen-Zee shift $\mathcal S$ dictating the response to underlying geometries~\cite{Haldane_sphere_1983,Wen:1992,gromov2015framing,bradlyn2015low}.
For Jain sequence $\nu = n/(2pn \pm 1)$ FQH states in \emph{abelian} topological order (we only consider abelian topological order throughout this work), the shift can be directly obtained from the $K$-matrix theory \cite{Wen:1992}, reading $\mathcal S = 2p \pm n$.
Near $\nu=1/3$, we define the generalized filling factor $\nu = (N+2)/[2Q + 3(2p+1)]$ \cite{Jain:2018}.
And near $\nu=2/3$, we define $\nu = (N+2)/[2Q + 3(2p-1)]$.
These generalized filling factors allow $\nu$ continuously tunable while recovering the correct shift at those fractional fillings.

The disorder potential is taken to be a sum of $N_{\rm imp}$ impurities:
\begin{equation}
V_{\rm dis}(\pmb{r}_i) = - \sum_{a=1}^{N_{\rm imp}} z_a v(|\pmb{r}_i - \pmb{R}_a|),
\end{equation}
where $z_a \in \{\pm 1\}$ is the charge of the $a$-th impurity with position $\pmb{R}_a$ having the same radius $R$ as the electrons on the sphere. 
$N_{\rm imp}$ is adjustable, and the positions are randomly sampled.
We include equal number of positive and negative impurities for a charge-neutral background.
We model $v(r)$ by a Gaussian form: 
\begin{equation}
\label{eq:gaussian}
v(r) = \sigma^{-1} e^{-r^2/2\sigma^2},
\end{equation}
with $\sigma$ controlling its spatial range. For short-range disorder, $\sigma$ should be smaller than all other length scales in the system: $\sigma < a_B, \ell_B, n^{-1/2}$, where $a_B = \epsilon \hbar^2/me^2$ is the effective Bohr radius and $n^{-1/2}$ is the average distance between electrons.
We scale the disorder potential by a disorder strength $\kappa_{\rm imp}$, this form of $v(r)$ qualitatively represents a screened charge with a characteristic energy scale $\kappa_{\rm imp} e^2/\sigma$ for $r<\sigma$. If $\kappa_{\rm imp}=\kappa$, this is equal to the electron-electron Coulomb strength. We intentionally choose $\kappa_{\rm imp}/\kappa=0.01$ to avoid the Coulomb bound states near short-range impurities, which is absent in experiments. This mimics experiments where varying the magnetic field modifies the cyclotron energy but leaves the bare disorder and Coulomb potentials unchanged.

To treat LL mixing and disorder under a unified framework, we extend the Psiformer~\cite{psiformer} and DeepHall~\cite{YQian_FQH_NNVMC_2025} framework to disordered FQH systems by introducing a disorder-aware self-attention ansatz.
The many-body wave function in our NQS is formulated in real-space Jastrow-Slater form \cite{psiformer, YQian_FQH_NNVMC_2025},
\begin{equation}
\Psi_{\lambda}(\pmb{r};\pmb{R}) = e^{J_{\lambda}(\pmb{r})} \det[M_{\lambda}(\pmb{r};\pmb{R})]
\end{equation}
where $\pmb{r} = (\pmb{r}_1, \dots, \pmb{r}_N)$ and $\pmb{R} = (\pmb{R}_1, \dots, \pmb{R}_{N_{\rm imp}})$ denote the electron and impurity coordinates, respectively, and $\lambda$ denotes the full set of trainable parameters in the model.
$J_{\lambda}(\pmb{r})$ is the electron-electron Jastrow factor.
The Slater matrix is constructed from configuration-dependent orbitals, $[M_{\lambda}(\pmb{r};\pmb{R})]_{ij} = \varphi_j(\pmb{r}_i;\pmb{r},\pmb{R})$, where each orbital $j$ depends not only on the coordinate of electron $i$, but also on the full electronic and impurity configurations. 

To encode disorder, the input feature layer assigns to each electron token both its coordinate $\pmb{r}_i$ and impurity-relative geometric features with respect to all disorder centers $\{ \pmb{R}_a \}_{a=1}^{N_{\rm imp}}$. We use raw relative vectors, rather than the logarithmically rescaled form adopted in Psiformer, because on the sphere all interparticle distances are bounded.
This impurity-aware featurization allows the ansatz to resolve disorder-induced spatial inhomogeneity while retaining the ability of real-space NQS to capture LL mixing beyond LLL projection.
Many-body correlations are encoded through self-attention~\cite{attention, psiformer}, which provides a flexible and permutation-equivariant representation of the electronic state.
Compared with electron-coordinate-only featurization, our input feature dimension is enlarged by a factor of $1+N_{\rm imp}$, increasing from $4N$ to approximately $4N(1+N_{\rm imp}$), while the token embedding dimension remains fixed.  
Training is performed in parallel on four NVIDIA H200 GPUs for each $\nu$ and $\kappa$. 
The variational parameters $\lambda$ are optimized within the VMC framework by minimizing the expectation value of the energy, with gradients computed by backpropagation through the neural-network wave function and the VMC estimator. 
Further implementation details and convergence tests are provided in the Supplementary Information.

{\em Phase Diagrams for $\nu < 1/2$}---Near $\nu=1/3$, we find that the liquid-solid competition is governed primarily by the FQHL and an e-WC, with increasing LL mixing and disorder favoring the latter.
Figure~\ref{fig:phase_diagram_results}(a-b) show the phase diagrams in the $\kappa$-$\nu$ plane near $\nu=1/3$, obtained for $N=12$ electrons. 
We fix $\kappa_{\rm imp}/\kappa = 0.01$ so that the short-range impurity potential remains weak compared to Coulomb and does not produce single-particle bound states throughout the phase diagram.
The color scale in Fig.~\ref{fig:phase_diagram_results}(a-b) represents the crystalline order parameter for $N=12$ e-WC on a sphere, $\ln P_6$, extracted from the angular power spectrum $P_L = (2L + 1)^{-1} \sum_{M=-L}^{L} \abs{c_{L,M}}^2$, where $c_{L, M} = \int d\Omega \, \rho(\theta, \phi)  Y_{L, M}^*(\theta, \phi)$ are the spherical harmonic expansion coefficients of the electron density obtained from the NQS: $\rho(\theta, \phi) = |\Psi_{\lambda}|^2$.
Similar to Bragg peaks in planar geometry, peaks in $P_L$ contains the information of crystalline order of a lattice constant $2\pi R/L_{\rm peak}$ on a sphere.
For example, near $\nu = 1/3$ where $N=12$ electrons form a regular icosahedron on the sphere, the icosahedral symmetry gives a peak at $L=6$, as shown in Fig.~\ref{fig:phase_diagram_results}(e).
In the nearly clean limit [Fig.~\ref{fig:phase_diagram_results}(a)] with $N_{\rm imp}=2$ (the minimum number of impurities required to perturb the system and break rotational symmetry of the Hamiltonian), the FQHL is confined to a narrow window around $\nu=1/3$. Neighboring fillings are predominantly crystalline, especially at large LL mixing. At weaker LL mixing ($\kappa \lesssim 8$), the FQHL expands over a broader filling range, qualitatively consistent with previous DMC result \cite{Jain:2018}. 

As the number of impurities increases from $N_{\rm imp}=2$ to $100$ [Fig.~\ref{fig:phase_diagram_results}(b)], crystalline order is strengthened in the large-$\kappa$ regime, as reflected by the larger values of $\ln P_6$ in Fig.~\ref{fig:phase_diagram_results}(b) compared to Fig.~\ref{fig:phase_diagram_results}(a). 
The most pronounced effect of adding impurities is at $\nu=1/3$, where the FQHL weakens and evolves toward a WC at sufficiently large $\kappa$, as directly shown in the density profiles in Fig.~\ref{fig:phase_diagram_results}(g-h).
To understand these results and their connections to experiments, we note that varying $N_{\rm imp}$ at a fixed electron number mimics the experimental tuning of short-range impurity density. By increasing the number of short-range impurities, the system approaches a ``white noise'' limit analogous to the Anderson model~\cite{anderson1958}, where the disorder is characterized entirely by its fluctuation strength, $\kappa_{\rm imp}$. Because the Anderson localization transition is driven by $\kappa_{\rm imp}$, keeping this parameter intentionally small ensures that our phase diagram avoids the Anderson localized phase. Consequently, the effect of introducing more impurities is to further lower the energy of the WC, stabilizing the crystalline phase against the competing liquid (see Appendix).

\begin{figure}
    \centering
    \includegraphics[width=\linewidth]{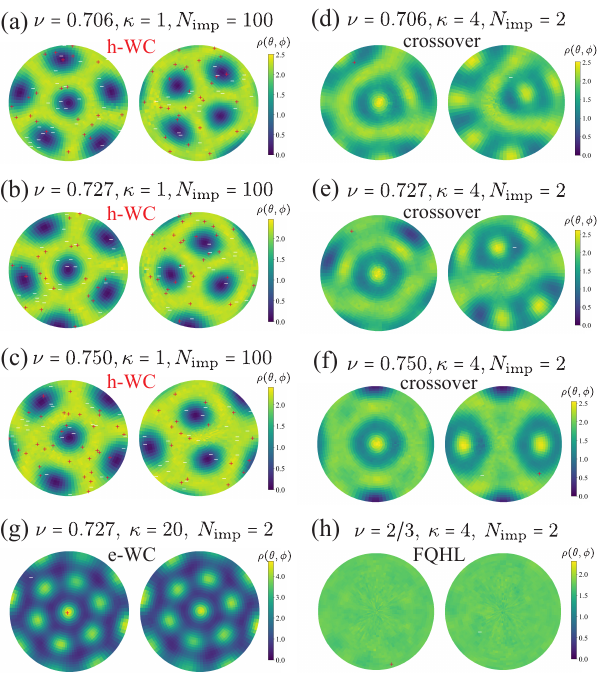}
    \caption{Electron density profiles projected onto the north and south hemispheres near $\nu=2/3$. 
    (a-c) h-WC states with 10, 9, and 8 holes, corresponding to $N_{\phi}$=31, 30, and 29, respectively.
    (d-f) Network-like density patterns in the crossover regime for different fillings.
    (g) A representative e-WC with $N=22$ electrons.
    (h) The FQHL at $\nu=2/3$. The $+$ and $-$ labels denote random impurities.}
    \label{fig:density_hemisphere}
\end{figure}

{\em Phase Diagrams for $\nu > 1/2$}---Above half filling, the phase diagram becomes richer. 
Near $\nu=2/3$, the liquid-solid competition involves not only the FQHL and an e-WC, but also a h-WC, as shown in Fig.~\ref{fig:phase_diagram_results}(c-d).
At exactly $\nu=2/3$, the FQHL remains robust over a broad range of $\kappa$, and crystallizes into an e-WC at large $\kappa$ only when impurities are added, similar to the behavior at $\nu=1/3$. 
Away from $\nu=2/3$, the key new feature is the presence of the h-WC stabilized at small $\kappa$, while an e-WC is favored at large $\kappa$. In electron density profiles, the h-WC appears as isolated ordered minima [Fig.~\ref{fig:density_hemisphere}(a-c)].
In our phase diagram, the h-WC appears for $\kappa \lesssim 1$, whereas the e-WC reenters at larger LL mixing: for $\kappa \gtrsim 12$ above $\nu=2/3$ and for $\kappa \gtrsim 4$ below $\nu=2/3$. 
The appearance of these distinct crystalline phases near $\nu=2/3$ can be intuitively understood from the two asymptotic limits. At small $\kappa$, near-exact particle-hole symmetry dictates that the low-density e-WC near $\nu=1/3$ must be mirrored by a h-WC near $\nu=2/3$.
At large $\kappa$, Coulomb repulsion dominates and stabilizes an e-WC. In this sense, LL mixing does not merely tune the competition between liquid and solid phases, but also selects between two distinct crystalline realizations of charge order.

We distinguish the two crystalline phases through their angular power spectra. 
For the e-WC, the number of lattice sites is fixed by the electron number $N$. For $N=22$, used in Fig.~\ref{fig:phase_diagram_results}(c-d), this gives a characteristic peak at $L_{\rm peak}=8$, so we take $\ln P_8$ as the e-WC order parameter. 
The h-WC, on the other hand, is determined by the hole number $N_h=2Q+1-N$~\cite{hole_number}, which varies with filling through the flux $2Q$. 
Near $\nu=2/3$, $N_h$ remains smaller than $N=22$, so the h-WC has a larger real-space lattice spacing than the e-WC and therefore a characteristic angular scale corresponding to $L_{\rm peak}<8$. At each filling we define the h-WC order parameter from the dominant peak, namely $\ln P_{L_{\rm peak} < 8}$, where $L_{\rm peak}$ is the value at which $P_L$ is maximal.

More remarkably, the evolution between these two crystalline phases is itself highly nontrivial. For fillings above $\nu=2/3$, we identify a novel crossover regime at $\kappa \sim 4-8$
[Fig.~\ref{fig:density_hemisphere}(d-f)] that connects the h-WC and e-WC.
The corresponding density profiles exhibit a network-like structure: the density is neither nearly uniform as in the FQHL, nor organized into isolated minima or maxima as in the h-WC or e-WC. Instead, the charge distribution forms a bicontinuous pattern, indicating that the h-WC-to-e-WC transition proceeds through a reconstruction of the density texture rather than a direct interchange between the two orders.
Moreover, the crossover regime is characterized by two distinct length scales, evidenced by a double-peak structure in the $P_L$ spectrum [see the $N_{\phi}=32, \kappa=4$ curve in Fig.~\ref{fig:phase_diagram_results}(f)]. 
Specifically, one peak occurs at $L=4$ and a second at $L=7$, with the latter closely matching the e-WC length scale. In addition, $P_L$ exhibits a deep minimum at $L=5$, the characteristic h-WC value of $L_{\rm peak}$ at the same filling. The quantity $\ln P_L$ evaluated at this same $L_{\rm peak}$ clearly separates the crossover regime from both the h-WC and e-WC phases, as shown in Fig.~\ref{fig:phase_diagram_results}(f), and is therefore used in the phase diagrams Fig.~\ref{fig:phase_diagram_results}(c,d).
This pronounced suppression of the $L=5$ signal implies that the h-WC melts into an intermediate phase within the crossover regime upon increasing $\kappa$.

In contrast, for fillings below $2/3$, the transition from h-WC to e-WC becomes significantly sharper, and within our $\kappa$ sampling we do not resolve a comparable crossover regime. 
This contrast reveals a pronounced asymmetry between the symmetry-broken phases on the two sides of the $\nu=2/3$ FQHL dome.
This asymmetry can be understood from two complementary viewpoints. 
Geometrically, the h-WC above $\nu=2/3$ is more dilute, so converting the widely separated density minima into the density maxima requires a more extended reconstruction of the density field. 
This favors a gradual percolation-driven reconstruction of the density texture and naturally gives rise to a broad network-like intermediate regime with increasing LL mixing. 
Below $\nu=2/3$, the hole density is higher and the closely packed density minima connect more easily, swiftly pinching off the interstitial electrons into isolated peaks and leading to a sharper transition into the e-WC.
Energetically, on the other hand, the denser electron background above $\nu=2/3$ can screen and soften the h-WC, making it more deformable under increasing LL mixing and thereby favoring a gradual reconstruction. 
Below $\nu=2/3$, this softening is weaker, so the reorganization into the e-WC is more abrupt.

{\em Discussion}---The central physical result of our study is the identification of a disorder-pinned h-WC at weak LL mixing near $\nu=2/3$, together with an intermediate-$\kappa$ crossover above $\nu=2/3$ that connects the h-WC to the e-WC. This h-WC provides a natural framework for understanding the reentrant IQH effect observed in GaAs experiments~\cite{Li_short_range:2010}, as well as its anomalous Hall analog in FCIs~\cite{Lu:2025,Xu:2025,Huang:2025}.
For the GaAs electron system~\cite{Li_short_range:2010}, $\kappa = 0.6$ at $B=15$T, corresponding to the center of the RIQH plateau near $\nu=0.63$ between $\nu=2/3$ and $\nu=3/5$. 
For twisted MoTe$_2$ at twist angle $\theta = 3.7^\circ$, the effective LL mixing is estimated to be $\kappa \sim 2$~\cite{mishra2026}, and a RIQAH plateau is observed in the hole-doped band over the same filling range~\cite{Xu:2025}. 
For pentalayer graphene on hBN, the effective LL mixing is estimated to be of order 2 to 6, with a RIQAH plateau also observed between $\nu=2/3$ and $\nu=3/5$~\cite{LL_mixing}.
In the limit of weak LL mixing ($\kappa \to 0$), the origin of h-WC follows from particle-hole symmetry: the e-WC near $\nu=1/3$ must be mirrored by a corresponding h-WC near $\nu=2/3$.
However, determining the stability window of such crystal phases in competition with the FQHL, and under realistic disorder and LL mixing, has remained a long-standing challenge~\cite{Maki1983, Girvin:1984, Levesque1984, Ortiz1993, Zhu1993, Zhu1995, Price1995, Kamilla1997, Yi1998, Yang2001, Mandal2003, Peterson2003, Archer2013}. 
Our NNVMC approach significantly advance this problem by using a unified, disorder-aware, and highly expressive neural-network ansatz capable of describing both the FQHL and WC simultaneously, and open a door towards future investigation of the crossover behavior between the two distinct symmetry-broken phases.
We note that our finite-size study specifically captures disorder-pinned crystal states, providing a direct connection to experimental observations without over-extrapolating to the true thermodynamic ground state of a clean system.

\section{Acknowledgments}
We thank Sankar Das Sarma and Long Ju for helpful discussions.
This work is primarily supported by the U.S. Department of Energy, Office of Basic Energy Sciences, under Contract No. DE-SC0025327. This research used resources of the National Energy Research Scientific Computing Center, a DOE Office of Science User Facility supported by the Office of Science of the U.S. Department of Energy under Contract No. DE-AC02-05CH11231 using NERSC award BES-ERCAP0032546.
This work was also facilitated through the use of advanced computational, storage, and networking infrastructure provided by the Hyak supercomputer system and funded by the University of Washington Molecular Engineering Materials Center at the University of Washington (DMR-2308979).

\appendix
\section{Appendix}
\begin{figure}[t]
    \centering
\includegraphics[width=1.0\linewidth]{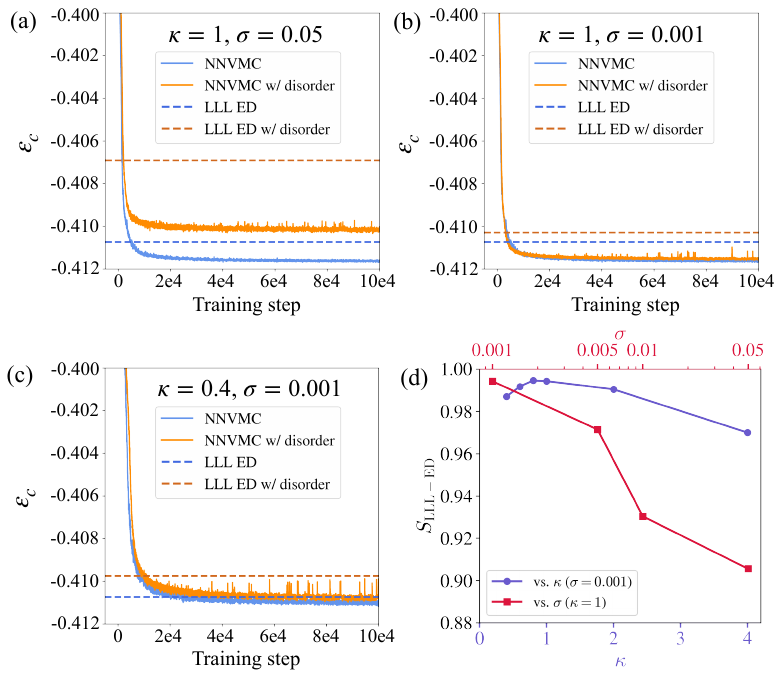}
\caption{ Energy and wavefunction overlap benchmark of NNVMC against LLL-ED for $N=8$ electrons at $\nu=1/3$ with a single repulsive impurity at the north pole. 
(a-c) Energy per electron $\varepsilon_c$ versus training step calculated by NNVMC (solid curves), and the ground state energy calculated by LLL-ED (dashed lines). Blue curves are for clean systems, without any disorder. Orange curves are for disordered systems with a single repulsive impurity placed at the north pole. In NNVMC, the impurity is modeled by a gaussian-type potential Eq.~(\ref{eq:gaussian}) with its length-scale $\sigma$ shown in each subtitle. In LLL-ED, the impurity is modeled by a delta-function.
(d) Wavefunction overlap with the LLL-ED versus $\kappa$ for a fixed $\sigma=0.001$ (blue) and versus $\sigma$ for a fixed $\kappa=1$ (red). $S_{\rm LLL-ED}$ is calculated at a fixed number of training step.
We fix $\kappa_{\rm imp} = 1$ in the NNVMC calculations here.
}
\label{fig:benchmark}
\end{figure}

{\em Benchmarking against ED}---We benchmark our NNVMC approach against LLL-projected ED (LLL-ED) by studying the $\nu=1/3$ FQH state. 
In the clean case without any impurity, the energy per electron $\varepsilon_c$ obtained from NNVMC during training is compared with the corresponding ED ground-state energy in Fig.~\ref{fig:benchmark}(a-c) shown as blue lines. Here $\varepsilon_c$ is corrected by taking into account the background charge and rescaled by $\sqrt{2Q\nu/N}$ to account for the density deviation from the thermodynamic limit intrinsic to spherical geometry.
For small $\kappa$, the NNVMC energy converges to the LLL-ED result [cf. Fig.~\ref{fig:benchmark}(c) blue lines], as expected in the $\kappa \to 0$ limit where the LLL projection becomes asymptotically exact.
At larger $\kappa$, the NNVMC energy converges below the LLL-ED value [Fig.~\ref{fig:benchmark}(a-b) blue lines], reflecting the fact that LL mixing lowers the energy beyond what can be captured within a strictly projected LLL treatment.

To benchmark the disorder response, we next consider a single impurity. In ED, the impurity is modeled as a delta-function potential projected onto the LLL, $V_{\rm imp}(\vb{r},\vb{r}_0) = V_0 \delta^2(\vb{r} - \vb{r}_0)$.
When the impurity is placed at the north or south pole, the Hamiltonian remains invariant under rotations about the $z$-axis, so $L_z=m$ remains a good quantum number. 
In this case, the impurity shifts only the energy of the $m=Q$ ($m=-Q$) orbital at the north (south) pole.
In the NNVMC calculation, the impurity is represented by the Gaussian potential with width $\sigma$ introduced in Eq.~(\ref{eq:gaussian}), whose $\sigma \to 0$ limit approaches the delta-function form used in ED.
To compare the two descriptions consistently, we match the impurity strength through $\kappa_{\rm imp}/\kappa = V_0/(2\pi\sigma)$ in the presented result shown in Fig.~\ref{fig:benchmark}. 
The corresponding energies with a single repulsive impurity placed at the north pole are shown in Fig.~\ref{fig:benchmark}(a-c) as orange lines.
As expected, the agreement with ED improves as $\sigma$ is reduced [Fig.~\ref{fig:benchmark}(a) and (b)], since the Gaussian impurity more closely approximates the delta-function potential, and also as $\kappa$ is reduced [Fig.~\ref{fig:benchmark}(b) and (c)], since LL mixing becomes less important.

We further assess the quality of the NQS variational wave function, especially when disorder is present, through its overlap with the ED wave function.
In the clean case, the overlap reaches $S_{\rm LLL-ED} \approx 0.995$ for $N=8$. 
With a single repulsive impurity, Fig.~\ref{fig:benchmark}(d) shows the overlap as a function of $\sigma$ at fixed $\kappa=1$ and as a function of $\kappa$ at fixed $\sigma=0.001$. 
In both cases, the overlap increases as the calculation approaches the projected-LLL limit with delta-function impurity, either by reducing $\sigma$ or by decreasing $\kappa$, and exceeds $0.99$ for sufficiently small $\sigma$ and $\kappa$. These results demonstrate that the NQS ansatz faithfully captures the physics of the FQH state, both in the clean limit and in the presence of weak disorder.
It should be noted that in Fig.~\ref{fig:benchmark}(d), $S_{\rm LLL-ED}$ does not increase monotonically as $\kappa$ decreases. Instead, for $\kappa < 1$, the overlap drops as approaching the LLL limit ($\kappa \rightarrow 0$). This is partially because the neural network requires more steps to converge for smaller $\kappa$, and Fig.~\ref{fig:benchmark}(d) is calculated at a fixed number of training steps. More fundamentally, reaching the $\kappa \rightarrow 0$ limit requires the real-space wavefunction to become strictly holomorphic to remain entirely within the LLL. Because our NQS operates on continuous coordinates without a rigid analytical constraint, any microscopic residual dependence on the anti-holomorphic coordinates ($\bar{z}$) is massively penalized by the kinetic energy operator. This amplifies the local-energy variance in VMC, forcing the optimizer to expend its capacity on suppressing kinetic energy fluctuations rather than resolving the delicate many-body correlations.

{\em WCs stabilized by short-range impurities}---We show that WCs can be stablized by short-range disorder against the uniform FQHL. 
We first recall the zero-field result~\cite{Joy:2025} and then extend the argument to the strong-field regime relevant for $\nu < 1$. To our knowledge, this strong-field result is presented here for the first time.

At zero magnetic field and sufficiently low density such that the kinetic energy $E_F \sim \hbar^2 n/m$ is much smaller than the Coulomb interaction energy $E_{\rm Coulomb} \sim e^2 n^{1/2}/\epsilon$ (i.e., $r_s = E_{\rm Coulomb}/E_F \gg 1$), electrons form a WC where each electron is confined by the Coulomb repulsion from other electrons. 
If we expand near the equilibrium position of a WC lattice site, we obtain a harmonic oscillator potential $m \omega^2 x^2\sim (e^2/2\epsilon r_{\rm avg}^3)x^2$, where $r_{\rm avg}$ is the average distance between electrons defined as $n \pi r_{\rm avg}^2 = 1$. 
This implies that the size of an electron wavefunction is $a\sim \sqrt{\hbar / m \omega} = (r_{\rm avg} a_B^3)^{1/4}$, where $a_B=\epsilon \hbar^2/me^2$ is the effective Bohr radius.
A description of the WC is justified if the size of the wavepacket is much smaller than the average distance between electrons, $a \ll r_{\rm avg}$ (equivalently, $a_B \ll r_{\rm avg}$ or $r_s = r_{\rm avg}/a_B \gg 1$).

Now we consider the effect of short-range impurities with delta-function potential $\sum_i V_i(\vb{r})=\sum_i V_0 \delta(\vb{r} - \vb{r}_i)$.
The short-range assumption is justified if its characteristic length scale $\sigma$ is much smaller than all other length scales in the system: $\sigma \ll a_B,r_{\rm avg}$.
By modeling the WC electron wavepacket as a Gaussian wavepacket $\rho(\vb{r}) = e^{-r^2/a^2}/(\pi a^2)$, we obtain the potential energy for the $i$-th impurity $V_i = V_0 \rho(\vb{r}_i)$ and its force $F_i = \grad V|_{r_i} = V_0 \grad \rho|_{r_i}$.
The mean-square-fluctuation of the force is given by
\begin{align}
    \ev{\Delta F^2} &= \int n_i dr^2 V_0^2 (\grad \rho \cdot \grad \rho), \\
    &=V_0^2 2\pi \int_0^{\infty} n_i r dr \frac{1}{(\pi a^2)^2} e^{-2r^2/a^2} \frac{4r^2}{a^4}, \\
    &= \frac{n_i V_0^2}{\pi a^4}.
\end{align}
As a result, the energy gain can be estimated by the average work done by the force $\Delta F$ countering the harmonic oscillator force $\sim m \omega^2 \Delta r$ with a displacement $\Delta r = \Delta F/m \omega^2$:
\begin{align}\label{eq:WC_energy_gain_0B}
    \Delta E_{\rm WC} = -\Delta F \Delta r = -\frac{n_i V_0^2 m}{2\pi \hbar^2}.
\end{align}

Similarly, we can estimate the energy gain for a uniform Fermi liquid (FL).
Consider a single short-range impurity.
The impurity potential induces a local density variation $\delta n$, which contributes to an increase in the kinetic energy $(\hbar^2 \delta n / m) \delta n r^2$ and the Coulomb self energy $(e^2/r) (\delta n r^2)^2$, and an energy gain $-V_0 \delta n$.
Optimizing the energy gives 
\begin{align}
    \delta n = \frac{V_0}{r^4(\frac{\hbar^2}{m r^2} + \frac{e^2}{r})},
\end{align}
and the corresponding energy gain in response to this single impurity reads
\begin{align}
    \delta E_{\rm FL,1} = -\frac{V_0^2}{r^4(\frac{\hbar^2}{m r^2} + \frac{e^2}{r})}.
\end{align}
$r\sim k_F^{-1} \sim n^{-1/2}$ corresponds to the smallest length scale of the density perturbation a FL can produce. 
As a result,
\begin{align}\label{eq:single_impurity_FL}
    \delta E_{\rm FL,1} = -\frac{V_0^2 n^2}{(\frac{\hbar^2n}{m} + e^2n^{1/2})} = - \frac{V_0^2 n m }{\hbar^2 (1+ r_s)}.
\end{align}
Equation~\eqref{eq:single_impurity_FL} is the total energy gain for all electrons in the presence of a single impurity.
In the presence of many impurities with concentration $n_i$, the average energy gain per electrons is 
\begin{align}\label{eq:Fermi_liquid_energy_gain}
    \Delta E_{\rm FL} = \frac{n_i}{n} \delta E_{\rm FL,1} = - \frac{V_0^2 n_i m }{\hbar^2 (1+ r_s)}
\end{align}
If $r_s \gg 1$, the parenthesis in the denominator of Eq.~\eqref{eq:Fermi_liquid_energy_gain} is dominated by $r_s$.
Comparing Eqs.~\eqref{eq:WC_energy_gain_0B} and \eqref{eq:Fermi_liquid_energy_gain}, we find that the energy gain of a WC is larger than the energy gain of a uniform FQHL by a factor of $r_s$.
Intuitively this can be understood that the Coulomb self energy of FL due to density variation in response to the impurity is comparable to the impurity potential energy gain by occupying/depleting the impurity position for negative/positive $V_0$.
As a result, the total energy gain of FL is a second order effect in $r_s$.
On the other hand, since there is no such Coulomb self energy to be overcome for the WC phase (the net effect is that some electron wavepackets shift their center, while the density profile of wavepacket themselves do not change), the total energy gain of WC is the not the second order effect, and being a factor of $r_s$ larger than FL. 

The above calculations are for zero-field case. 
Now we consider the strong-field case.
If $B$ is sufficiently large such that $l_B < a < r_{\rm avg}$, the size of a wave packet is replaced by $l_B$ .
For example, in GaAs 2DES, take $n\sim 10^{11}$ cm$^{-2}$, $a_B = 10$ nm, then such a criterion gives $B>1.2$ T.
The density of the wavepacket now is described as
\begin{align}
    \rho(\vb{r}) = \frac{1}{2\pi l_B^2} e^{-r^2/2l_B^2},
\end{align}
which corresponds to a different force executed by the random impurities
\begin{align}
    \Delta F = \sqrt{\frac{n_i V_0^2}{\pi l_B^4}},
\end{align}
and a different displacement
\begin{align}
    \Delta r = \frac{\Delta F}{m\omega^2}.
\end{align}
As a result, the energy gain reads
\begin{align}
    \Delta E_{\rm WC} = -\Delta F \Delta r = -\qty(\frac{r_{\rm avg}}{l_B})^4 \frac{n_i V_0^2}{r_{\rm avg} e^2}.
\end{align}
For a uniform FQHL, there is no kinetic energy and by replacing the smallest length scale by $l_B$ we obtain
\begin{align}
    \Delta E_{\rm FQHL} = - \qty(\frac{r_{\rm avg}}{l_B})^3 \frac{n_i V_0^2}{r_{\rm avg} e^2}.
\end{align}

Indeed, the energy gain of the WC is greater than the energy gain of the FQHL by a factor of $r_{\rm avg}/l_B = 1/\sqrt{\nu}$.
As a result, when the bare energy difference (i.e., the energy calculated without disorder) between WC and FQHL is not very large, often within 1\% for numerical calculations~\cite{Zuo2020,Jain:2025}, adding disorder can dramatically shift the phase boundary between WC and FQHL and stabilize WC especially for small $\nu < 1$.


\begin{thebibliography}{64}
\bibitem [{LL_()}]{LL_mixing}%
  \BibitemOpen
  \href@noop {} {}\bibinfo {note} {We estimate the effective LL mixing in pentalayer graphene/hBN as $\kappa_{\rm eff} \sim e^2/\epsilon a_{\rm eff} \Delta_{\rm iso}$, with $\epsilon \approx 5$, $a_{\rm eff} \approx 3-6$ nm, and $\Delta_{\rm iso} \approx 10-20$ meV. The range of $a_{\rm aff}$ is inferred from the spatial extent of the active Chern band orbital for a moir\'e scale of roughly $12-15$ nm, while $\Delta_{\rm iso}$ is taken to be of the same order as the first conduction-band bandwidth, which can be as small as $\sim 10$ meV. These values give $\kappa_{\rm eff} \sim 2-6$.}\BibitemShut {Stop}%
\end{thebibliography}

%

\end{document}